# V5856 Sagittarii/2016: Broad Multi-Epoch Spectral Coverage of a Sustained High Luminosity Nova

Robert Williams[1,2], Frederick M. Walter[3], Richard J. Rudy[4,7], Ulisse Munari[5], Paul Luckas[6], John P. Subasavage[7], and Jon C. Mauerhan[7]

[1]*Space Telescope Science Institute, 3700 San Martin Drive, Baltimore, MD 21218, USA; wms@stsci.edu*
[2]*Department of Astronomy & Astrophysics, University of California, Santa Cruz, 1156 High Street, Santa Cruz, CA 95064, USA*
[3]*Department of Physics and Astronomy, Stony Brook University, Stony Brook, New York, 11794, USA*
[4]*Kookoosint Scientific, 1530 Calle Portada, Camarillo, CA, 93010, USA*
[5]*INAF, Osservatorio Astronomico di Padova, 36012 Asiago (VI), Italy*
[6]*International Centre for Radio Astronomy Research, The University of Western Australia, 35 Stirling Hwy, Crawley, Western Australia 6009*
[7]*Physical Sciences Laboratory, The Aerospace Corporation, M2-266, P.O. Box 92957, Los Angeles, CA 90009*

## Abstract

Nova V5856 Sagittarii is unique for having remained more than nine magnitudes above its pre-outburst brightness for more than six years. Extensive visible and IR spectra from the time of outburst to the present epoch reveal separate emitting regions with distinct spectral characteristics. Permitted emission lines have both broad and narrow components, whereas the forbidden line profiles are almost entirely broad. The permitted line components frequently display P Cygni profiles indicating high optical depth, whereas the broad components do not show detectable absorption. The densities and velocities deduced from the spectra, including differences in the O I λ7773 and λ8446 lines, are not consistent with an on-going wind. Instead, the prolonged high luminosity and spectral characteristics are indicative of a post-outburst common envelope that enshrouds the binary, and is likely the primary source of the visible and IR emission.

## 1. Introduction

V5856 Sagittarii (ASASSN-16ma; N Sgr 2016d; hereafter V5856 Sgr) was the fourth nova detected in outburst in the constellation Sagittarius in 2016. It was discovered 2016 October 25.02 by Stanek et al. (2016) and shown to be in the Fe II spectral phase a few days later by Luckas (2016). Early interest in V5856 Sgr emerged when it was discovered to be a bright γ-ray source. The nova was detected by the Large Area Telescope (LAT) on *Fermi* Observatory in γ-rays over the 10-day period 2016 November 7-17 (Li et al. 2016). Observed in only a dozen or so novae in the past decade, γ-rays are usually detected shortly after outburst for only a period of a few days, and not with the luminosity of V5856 Sgr. At this time Munari, Hambsch, and Frigo (2017) presented detailed optical photometry of the nova that showed a clear correlation between the γ-ray brightness and the optical brightness, a correlation that was previously explored by Li et al. (2017) and Aydi et al. (2020).

When the nova's sustained high brightness became evident, UM and his group at Asiago observatory included V5856 Sgr in their nova follow-up program, obtaining both photometry and occasional spectra. They focused on determining its photometric characteristics, with the results of their analyses reported by Munari et al. (2021, 2022). To illustrate its photometric behavior we show the visible light curve of V5856 Sgr in Figure 1, as obtained from various sources, including the AAVSO website[1]. As recounted by Munari and colleagues, the nova does exhibit photometric characteristics that are unusual for novae: (1) The rise to visual maximum brightness was steady and relatively slow, requiring two weeks to arrive at its peak, and (2) the nova's decline in brightness, also gradual, flattened out six months after outburst at an unusually high brightness level----after six years remaining more than 9 magnitudes above its pre-outburst brightness. This sustained brightness is rare, an exception to the large majority of novae that return to pre-outburst brightness within a few years.

It Is the nova's prolonged brightness and spectral evolution that are the primary focus of this paper—a topic that is addressed through the presentation and analysis of an extensive set of visible and infrared spectroscopic data.

## 2. Early Decline Spectrum

V5856 Sgr was not observed extensively during its period of early decline. Even after the *Fermi* detection of high energy γ-rays six weeks after outburst, the nova did not attract a great deal of attention from observers. The only spectra of which we are aware taken near the time of outburst are a series of seven low resolution visible spectra obtained by PJL of the world-wide ARAS group of astronomers. Taken between 2016 October 27---November 18 following its discovery on 2016 October 25, the spectra are posted on the ARAS website[2] for public download. Over this 3-week period, clear changes were observed in the spectra, in several cases occurring rapidly over a period of 24 hours. The spectra, with a resolution R=530 (equivalent to 9 Å at Hβ, or 550 km/s), are displayed in Figure 2.

The initial spectrum on 27 October shows an Fe II-phase spectrum, with the Balmer lines and Fe II multiplet 42 appearing in emission. Within three days the initial spectrum changed so that, with the exception of Hα, the prominent features had all converted to absorption. The spectrum maintained basically the same form for two weeks, during which time numerous non-Balmer absorption lines steadily become more prominent. In this phase the majority of the stronger absorption lines were Fe II, Na I, and Si II transitions with a sample of other low ionization lines arising from an effective photosphere that had a velocity dispersion of ~1,000 km/s based on observed line widths. The ejecta had an expansion velocity of a lower value from the absorption displacements of the P Cygni profiles.

We show in Figure 3 the spectrum on 9 November, the day after maximum visible brightness when the absorption lines were most prominent, together with line identifications we have made for which we have some confidence. The relatively low resolution of the spectrum causes sufficient line blending that some of the line identifications remain uncertain. The absorption lines may belong to a transient heavy element absorption (*thea*) system (Williams et al. 2008) that would be revealed in more detail at higher spectral resolution.

---

[1] https://www.aavso.org/LCGv2/
[2] https://aras-database.github.io/database/novasgr2016d.html

By 12 November the stronger absorption lines, e.g., the Balmer and Fe II transitions, again developed emission components. Within a week, on 18 November, the prominent absorption lines had fully transformed back into emission lines, indicating the development of a prominent emission region above the photospheric continuum region. At this time absorption lines were hardly detectable in the continuum and the spectrum again resembled the initial early post-outburst Fe II phase spectrum observed on 27 October, two days after discovery, but with broader emission components indicating a higher velocity of ~1,000 km/s for the emitting ejecta.

### 3. Subsequent Spectroscopic Development

#### 3.1. Infrared Spectrum

Following the early decline in brightness of the nova and the cessation of its 10-day period of gamma ray detection, little spectroscopic attention was given to V5856 Sgr while the star was in conjunction with the sun, from mid-November to mid-February. As part of their ongoing IR studies of novae a group at Aerospace Corp led by RR obtained visible and IR spectra on an annual basis. The infrared spectra were acquired with the Aerospace Corporation's Visible and Near-Infrared Imaging Spectrograph (VNIRIS) described by Rudy et al. (2021), mostly using the 3m Shane telescope of Lick Observatory. A collection of their IR spectra, with a resolution of 10 Å at 1 micron (300 km/s), are shown in Figure 4. The initial spectrum was obtained on 2016 November 23, almost coincident with the final ARAS optical spectrum, and identifications are shown for the emission lines. Other than a decline in the equivalent widths of the emission lines, i.e., their strengths relative to the continuum, the evolution of the IR spectrum does not show significant changes from the first IR spectrum obtained one month after outburst until the present time.

Since the time of the earliest spectra the J band, shortward of 1.3 microns, has displayed a diverse group of neutral species of H, He, C, N, and O, that include the forbidden line [N I] $\lambda 1.0405$ $\mu$ one month after outburst. The H band is dominated by lines of the H I Brackett series, with another broad forbidden line [Fe II] $\lambda 1.677$ $\mu$ also present. The K band displays two strong lines, He I $\lambda 2.058$ and H I Br-$\gamma$ $\lambda 2.1655$ microns, with weaker features at $\lambda 2.12$ and $\lambda 2.207$ microns also present in 2016, probably due to C I and Na I, respectively (Das et al. 2008). The clear presence of C I and Na I lines, together with the weaker Mg I $\lambda 1.5749$ $\mu$ line blended with the red wing of the Br15 line in the 2016 November 23 spectrum is notable.

Beginning with the early detection of emission of the first overtone $\Delta v=2$ carbon monoxide (CO) band at 2.29 $\mu$ by Evans et al. (1996) for the nova V705 Cas/1993, which subsequently formed dust, a relationship has emerged between CO and neutral metal line emission. In the years following the Evans et al. observations, additional novae were observed with carbon monoxide IR band emission shortly after outburst that also formed dust. In their study of nova V2274 Cyg/2001, Rudy et al. (2003) noted the strength of the correlation between dust formation and CO emission in novae. Subsequently, Das et al. (2008) extended this correlation to include the strength of low ionization emission species Na I and Mg I as reliable indicators of cool gas that favored subsequent dust formation. In a detailed study of nova V496 Scuti/2009, Raj et al. (2012) observed CO emission early in decline, followed five months later by clear dust formation. Noting the strong presence of neutral C I, Na I, and Mg I emission lines in the spectrum, they found from modeling the ejecta that detectable CO band emission is expected in conditions when those neutral metal lines are strong.

The extent to which V5856 Sgr follows this paradigm is unclear. The 2016 November 23 spectrum shown in Fig. 4 does show C I, Na I, and Mg I lines, where the C I lines are observed to be strong, but not the Na I and Mg I lines. The long wavelength cutoff of the spectrum occurs at 2.30 $\mu$, just past the limit of the first overtone CO band at 2.29 $\mu$, for which there is no clear detection. Assuming that a brief appearance of CO emission was not missed before or after our IR observations, V5856 Sgr may be an example where neutral metal lines in the JHK band region are not necessarily indicators of detectable CO band emission unless the Na I and Mg I are rather stronger relative to the H I and He I lines than we have observed. That is, strong C I, Na I, and Mg I emission lines could be a necessary, but not sufficient, condition for the presence of detectable IR CO overtone band emission in novae.

The IR spectra clearly show low ionization species to be present during the entire period of decline, including the current elevated luminosity phase. The H and He lines originate from direct electron-ion recombination, and the CNO emission features longward of 1 micron are also likely to have a substantial fraction of their excitation from direct 2-body electron recombination. However, there is some evidence against recombination being the only process producing the neutral CNO lines in the IR because the observed strengths of the IR lines are much greater than the strongest optical lines expected for each ion from 2-body recombination (Nussbaumer & Storey 1984; Pequignot, Petitjean, & Boisson 1991). Some of this difference is due to reddening in the emission region (Munari et al. 2022). However, in the same way that resonance fluorescence of H Ly-$\beta$ emission by the O I $\lambda$1027 resonance line produces strong O I $\lambda$8446 emission in most novae (Bowen 1947), it is likely that a fraction of the observed neutral CNO emission features may be produced by a process such as resonance fluorescence of continuum radiation followed by collisional transfer between excited states (Williams & Ferguson 1982; Rudy et al. 1991), rather than by direct recombination.

### 3.2. *Visible Spectrum*

Following the initial IR spectrum, no subsequent spectra were obtained of V5856 Sgr for a period of seven months until 2017 June 18-19, when FW obtained a pair of low dispersion spectra using the Goodman High Throughput Spectrograph (GHTS) on the SOAR 4m telescope. One spectrum covered the 3100-5800 Å region with a resolution of ~2000 using the 400 l/mm grating + UV filter; the other covered the 5600-6800 Å region at a resolution of ~5500 using the 1200 l/mm grating. These spectra were dominated by H I Balmer lines, He I, and assorted forbidden lines. The higher resolution red spectrum showed the forbidden lines to be broad and flat-topped. H$\alpha$ and He I $\lambda$5876 had narrow emission cores atop broad bases having widths of 1800 and 1500 km/s, respectively. The blue spectrum was complex and better deconvolved in subsequent higher resolution Chiron spectra that commenced one month later.

From 2017 July – 2022 July FW added the nova to his program of acquiring high-resolution spectra of novae with the CTIO/SMARTS 1.5-m telescope that would enable reliable line identifications, profiles, and radial velocities to be determined, as shown by Munari et al. (2022). During this time a series of 26 spectra were obtained with the Chiron fiber-fed bench mounted echelle spectrograph at spectral resolution R=28,000, or 11 km/s (Tokovinin et al. 2013). Generally taken at weekly to monthly intervals, the total exposure times ranged from 15 to 50 minutes, typically in coadded 15-20 minute integrations. The data were reduced using a pipeline coded in IDL[3]. The images were flat-fielded, and cosmic rays were removed using the L.A. Cosmic algorithm (van Dokkum 2001). The 74 echelle orders were extracted using a boxcar extraction, and instrumental background computed on both sides of the spectral trace was subtracted. These publicly accessible spectra[4] constitute an excellent archive of the progressive changes in the spectrum of a post-outburst nova at a resolution that enables reliable line profiles and radial velocities of the features to be determined.

---
[3] http://www.astro.sunysb.edu/fwalter/SMARTS/CHIRON/ch\_reduce.pdf

As Chiron is fiber-fed, there is no simple method to subtract the sky. The fibers have a diameter of 2.7 arcsec on the sky. Given the brightness of V5856 Sgr night sky emission was generally negligible apart from narrow [O I] and Na D lines, and some OH airglow lines at longer wavelengths. Wavelength calibration used ThAr calibration lamp exposures at the start and end of each night. The instrumental response was removed from the individual orders by dividing the spectrum by that of a flux-standard star, μ Col. This provided flux-calibrated orders with a systemic uncertainty due to sky conditions. Individual orders were spliced together, resulting in a calibrated spectrum from 4083-8900 Å with 5 inter-order gaps in the coverage longward of 8260 Å. Contemporaneous BVRI photometry was used to scale the spectrum to approximately true fluxes.

More recently, UM has used the telescopes at Asiago Observatory to obtain spectra of V5856 Sgr at both high and low spectral resolutions. A montage of the spectra obtained on the SMARTS/CTIO, Aerospace Corp., and Asiago Observatory telescopes in the period 2016-2021 are shown in Figure 5, depicting how the spectrum has evolved at visible wavelengths during that time. The more prominent lines include (a) permitted recombination lines of H and He I, (b) collisionally excited forbidden lines that include [N II] λ5755, [O III] λλ5007,4959, and [O II] λλ7319/31, (c) CNO recombination lines observed at several epochs, that include the O II λ4651 multiplet, and more weakly C II λ4267, and (d) O I λ8446 and λ7773, that are populated by resonance fluorescence of H Ly-β (Bowen 1947) with possible collisional transfer between the O I triplet and quintet states (Kastner & Bhatia 1995). In addition to these lines (e) there are present in many spectra broad N III λ4640 (mult 2), and in a few spectra C IV λ5805, both excited by UV fluorescence scattering by the strong principle resonance lines N III λ374 and C IV λ312, with which they share the same upper levels.

The numerous high-resolution SMARTS/Chiron spectra are of particular importance in deciphering the characteristics of the ejecta from line profiles. We have chosen a selection of emission lines in the visible spectrum that are populated by different excitation processes, and show them individually in Figures 6-11. The spectra are displayed at normalized continuum flux levels because of some observations made on nights that were not necessarily photometric. The line profiles are well resolved and show kinematically distinct emission regions of the ejecta.

With few exceptions the quality of all the visible and IR spectra are good, with only a few of the approximately 40 spectra taken through thin cirrus. Except for the ARAS spectra obtained immediately after outburst in late October/early November, for which only one observation was made at an airmass greater than 2.0, all spectra but two were taken in the interval March-August and at airmass less than 1.5.

We have used the high-resolution line profiles to deconstruct the emission lines into separate velocity components, attempting to account for the largest fraction of flux with a minimum of distinct components. In defining components having different velocity structures we confined our attention to optically thin lines that had little appreciable absorption detected. Visual inspection of the profiles displayed in Figs. 6-11 indicates that the majority of lines can be fit reasonably well with two principal velocity components---one broad, one narrow. The permitted transitions Hβ, He I λ6678, and O I λ8446 have line profiles that can be fitted well with a combination of a broad trapezoidal feature as a base, upon which is superposed a narrow Gaussian-shaped feature, where the relative strengths of the two components vary over time. The broad component is strongest shortly after outburst and systematically weakens in time with respect to the narrower feature, that becomes more dominant in recent years. The forbidden lines show only the broad, flat-topped 'boxy' profile whose flat peak does exhibit intensity variations. No narrow feature is present in the forbidden lines with the exception of the auroral transition [O III] λ4363, which does exhibit a weak narrow component in several spectra.

---

[4] http://www.astro.sunysb.edu/fwalter/SMARTS/NovaAtlas/v5856sgr/v5856sgr.html

In Figure 12 we show the deconvolution of several of the stronger lines into separate narrow and broad components at one typical epoch for 2018 March during decline. An overplot of the permitted and forbidden lines together in velocity space, shown in Figure 13, demonstrates the broad profiles of the permitted and forbidden lines to be quite similar. The broad components of both types of lines have widths with FWHM=635 km/s, with small differences that indicate they may be formed in the same spatial region, although likely at different densities because of their different critical densities. The narrow components of the permitted lines have widths with FWHM=50 km/s.

The following features stand out upon examination of the Chiron nova spectra observed at a resolution of 11 km/s:

(i) The permitted H and He I lines consist of a narrow component superposed on a much broader base. Deconvolution of the profiles shows the narrow component to be approximately Gaussian and have a FWHM=50 ± 3 km/s. The broad component width has FWHM=635 ± 21 km/s. It is significant that both the narrow and the broad components of the H and He lines show no measurable changes in width or radial velocity over the 5-year period of observations apart from small changes in the line profiles between spectra.

(ii) The forbidden lines are broad and flat-topped, i.e., roughly rectangular, over the entire time interval with widths FWHM≈635 km/s, and they lack a detectable narrow component. [O III] λ4363, whose critical density of ~$10^8$ cm$^{-3}$ is highest among the forbidden lines, is an exception in exhibiting in a few spectra a weak, narrow emission component superposed on its broad component. Initially observed in 2017 July, although likely present earlier, [N II] λ5755 first appears stronger than [O III] λ5007, as is expected at higher densities due to the higher critical density of the auroral [N II] line. With time, however, [O III] becomes stronger than [N II] from an increasing ionizing radiation field. The forbidden lines do show minor changes in their structure, but no reliable radial velocity variations are measured for the forbidden lines over time. Notably, the forbidden lines weakened dramatically during 2019, during which time O I λ7773 changed from an emission to an absorption feature. The forbidden lines regained strength In 2020.

(iii) The narrow components of Hα, Hβ and He I λ6678 frequently show P Cygni type absorption, indicating significant optical depth. Hβ absorption is usually present in most spectra, although normally weak compared to its much stronger emission component. The He I line presents absorption on occasion, with the strengths of the absorption and emission components varying widely. Usually, the He I emission dominates the absorption component, but there are epochs when the reverse is true. The broad components of these lines lack clear evidence for absorption in any of the spectra, but could be present at weak levels that are difficult to detect against the continuum.

(iv) O I λ8446, excited by fluorescence of Ly-β, has a strong narrow emission component at all epochs that O I λ7773 never exhibits. This is significant in constraining the densities where the two lines are formed. The absence of λ7773 emission in the narrow-line emitting region at fluxes comparable to those of λ8446 indicates that the narrow component of λ8446 must be formed at densities $\lesssim 10^{12}$ cm$^{-3}$ (Kastner & Bhatia 1995), otherwise at higher densities collisional transfer between the O I triplet and quintet states will produce an intensity of λ7773 comparable to that of λ8446, which is clearly greater than that observed. The λ7773 multiplet does at some epochs exhibit an absorption feature due to population of the metastable O I 3s $^5$S$^o$ level by processes that are not related to the λ8446 emission process. The width of the narrow component of O I λ8446 is the same as that of the Balmer lines.

(v) He II λ4686 and the C IV λ5805 $^2$S-$^2$P$^o$ doublet are the only higher ionization emission lines detected in the spectrum of V5856 Sgr. They appear weakly and have variable strengths between 2018-2022, with broad profiles. They lack any narrow component observed for the H and He I lines. The C IV line is very likely excited by fluorescence scattering of continuum radiation by the C IV λ312 Å resonance line, since the two lines share the same upper level.

(vi) Numerous CNO permitted lines appear in the nova spectra that are expected to be excited by electron recombination (Nussbaumer & Storey 1984; Pequignot et al. 1991). In the visible these lines include C II λ4267, C II λ7235, and N II λ5679, and they are observed as weak features having predominantly narrow components similar to those of the H and He I lines. It is possible that broad components for these lines exist, but are too weak to be detected against the continuum. Numerous CNO permitted lines appear in the infrared spectra, where they are much more prominent than those in the visible (cf. Figs 4 & 5). The relative intensities of most of the CNO lines deviate from those predicted by direct electron-ion recombination. Part of this is due to reddening with color excess of E(B–V)=0.32 (Munari et al. 2022) and partly due to the fluorescence scattering of continuum radiation by resonance lines that is contributing to the excitation of the CNO permitted transitions (Williams & Ferguson 1982).

## 4. Structure of the Emitting Region

The prominence of the very narrow emission features exhibited by most of the permitted lines is one of the unusual spectral characteristics of V5856 Sgr. The narrow components of H and He I have widths of FWHM≈50 km/s, and are frequently accompanied by clear P Cygni absorption features whose strengths vary with the optical depth of the line. Apart from small shifts in the emission wavelength caused by the varying strength of the absorption component, no appreciable radial velocity variations greater than ±8 km/s have been detected in the narrow components. Strong narrow emission components are observed in a small fraction of novae, with their widths varying considerably for different novae (Takeda & Diaz 2015). They are frequently interpreted as emission from the mid-section of a dumbbell structure characterizing early ejecta (Munari, Mason, & Valisa 2014) or from narrow conical-shaped jets ejected from the binary system (Shore et al. 2018). In some cases radial velocity variations are measured for the narrow lines, and some show a periodicity.

It has long been noted that novae forbidden emission lines tend to have a box-like shape (Menzel & Payne 1933) with widths that are similar to those of the Balmer lines (Williams & Mason 2010). Beals (1931), Chandrasekhar (1934), and Bappu and Menzel (1954) all performed calculations that show flat-topped, squarish line profiles are the result of emission by "a transparent expanding shell" of different thicknesses and velocities, where the flatness of the intensity peak and the sharpness of the drop in intensity are related to how thin the shell is relative to its radius. V5856 Sgr is typical in this regard in that the forbidden lines all have large widths that are the same as the broad components of H and He I, as shown in Fig. 13. Whatever their spatial distribution, the forbidden lines must originate preferentially in lower density gas with n$\lesssim 10^8$ cm$^{-3}$.

The fact that V5856 Sgr has remained luminous for some years after outburst is suggestive of continuing activity that is likely to produce ejection of mass via a wind or a more ballistic ejection of gas with discrete globs that expand and interact with each other (Aydi et al. 2020). If a wind, the flow will have higher density in the lower velocity inner regions. If ballistic ejection of globules is the dominant mechanism the density and velocity distribution will be more complex. Given the fact that numerous novae that have spatially resolved ejecta show pronounced inhomogeneities in a shell, e.g., GK Persei (Shara et al. 2012), T Pyxidis (Shara et al. 1997), and HR Del (Harman & O'Brien 2003), emission arising from a distribution of ejected globs that have density gradients is a possibility for the origin of the broad components of the forbidden and permitted lines in V5856 Sgr. Photoionization by UV, X-ray, and gamma radiation and/or collisional interaction of the globules among themselves or with ambient gas explains the ionization and excitation (Chomiuk, Metzger, & Shen 2021).

An alternative origin of the broad emission component is the possibility that with their large velocity widths the emission comes from a wind originating from the hot WD surface. The calculation of emission-line profiles in stellar winds has a rich history, with results showing that the optical depth of the gas and the expansion velocity law are key parameters that dictate the profiles (Castor & Lamers 1979). Realistic wind velocity laws expected for thermal Parker-type (Parker 1958) and general X-type supersonic winds with disks (Shu et al. 1994) produce emission lines with narrowly peaked profiles (Kudritzki et al. 1989; Hillier 1991). Winds do not generate the boxy, flat-topped emission profiles that are observed in V5856 Sgr, a fact that argues against the formation of the forbidden lines and the broad components of the permitted lines in a spherical outflowing wind. It is possible, however, to produce roughly flat-topped emission profiles from a bi-polar conical outflow with selected geometry and velocity law, augmented by emission from mass transfer between the secondary star and WD (Munari et al. 2011; Shore et al. 2018).

The origin of the prominent narrow emission component of the permitted lines, with widths of 50 km/s is uncertain. It was first observed seven months after outburst and has persisted since that time. The presence of P Cygni absorption associated with the narrow feature in some lines demonstrates that its emitting region is optically thick in those lines. For V5856 Sgr, whatever process produces the narrow component must continue for the six-year duration of our observations. Its small velocity width is indicative of an association with the secondary star rather than the more massive, compact WD, where much higher velocities are expected.

An important feature of the narrow emission component is its lack of radial velocity variations in excess of ±8 km/s. Unlike the broad, flat-topped forbidden lines, the narrow emission component could conceivably originate in continuous ejecta from the secondary that is driven by a wind from the WD, which should lead to emission from both the secondary ejecta and the WD wind. However, in order to explain the lack of radial velocity variations, this scenario is possible for V5856 Sgr only if the orbital plane happens to be closely aligned with the plane of the sky. A post-outburst mixture of gas from both stars that is optically thick and obscures the stars and their orbital motion is an equally likely explanation for the emission having no detectable radial velocity variations. If such a structure is stable it could also account for the prolonged high visible luminosity of the nova.

## 5. A Common Envelope?

A realistic explanation for the elevated luminosity and spectral characteristics of V5856 Sgr may exist in the form of a stable post-outburst common envelope (CE) for the binary.  Common envelopes in mass transfer binaries have long been advocated as normal configurations for nova systems, produced by heating of the Roche lobe-filling secondary from WD radiation and impact of its ejecta and wind (Paczynski 1976; Iben & Livio 1993).  As configurations of optically thick gas that are not necessarily in hydrostatic equilibrium, common envelopes receive significant energy from the embedded stars.  They are stable in the sense of being gravitationally bound to the binary, and normally with mass loss over time-scales much longer than the orbital period of the stars.  The spectra of novae at outburst almost always show the ejecta initially to be optically thick, either in the form of gas ballistically ejected (Shore 2014) or in the form of a wind driven from the hot WD surface (Aydi et al. 2020; Chomiuk et al. 2021).

   Normally, mass loss decreases soon after outburst and the rapid expansion of the nova ejecta drives down their optical depth to the point where they constitute a rapidly expanding emitting shell rather than a stationary common envelope.  The high luminosity of V5856 Sgr may be explained by such an envelope that is not transitory, but maintained for months or years by ongoing nuclear reactions on the WD and frictional heating from the orbiting stars within the envelope (Sparks & Sion 2021).  The fact that specific ejecta geometries, which can be so varied, are important in determining the general characteristics of novae spectra explains why a wide variety of novae light curves and spectra are observed.  Based on the similarity of V5856 Sgr's spectral evolution compared with other novae, e.g., it does exhibit a *thea* absorption system, there is good reason to consider it to be a classical nova whose slow rise to and fall from maximum light and its unusually narrow emission features are due to its geometry differing from that of most other novae.

   Specific conditions are required for a common envelope to be stable for prolonged post-outburst brightness and frictional heating to maintain the observed luminosity.  The momentum transferred by a compact object to a homogeneous medium through which it moves, gravitationally accreting gas as it converts its kinetic energy to thermal energy of the medium, was studied by Hoyle & Lyttleton (1939).  They showed the rate of accretion could be expressed as

$$dM_{acc}/dt = 4\pi\xi\rho_o G^2 M^2/v^3, \qquad (1)$$

where object of mass M moves through gas of density $\rho_o$ at velocity v with respect to the medium. The numerical factor $\xi \sim 1$ depends on several factors, e.g., whether the motion of the object is super- or sub-sonic in the medium (Bondi & Hoyle 1944).  The transfer rate of kinetic energy to the medium is approximately the multiplication of equation (1) by $v^2$.  If the thermal energy is largely radiated away it corresponds to the luminosity of the common envelope around embedded stars, which is related to their mass, relative velocity, and the local gas density.

   For V5856 Sgr the gas density that would be required to support a radiating common envelope can be determined approximately.  Munari et al. (2022) made ground-based IR and Gehrels *Swift* UV and visible observations from which they determined the post-outburst luminosity of V5856 Sgr to be $4 \times 10^3$ $L_\odot$.  We assume typical cataclysmic variable parameters for the nova, having a M=1 $M_\odot$ WD and a secondary star mass of 0.5 $M_\odot$ orbiting with a velocity of 250 km/s.  Ignoring the motion of the more massive WD, and assuming for the moment a non-rotating envelope, the gas number density required for interaction between the secondary star and the envelope to account for V5856 Sgr's observed luminosity is $n_o \sim 3 \times 10^{15}$ cm$^{-3}$.  This density pertains to the region of the envelope near the secondary star, as is determined by the mass loss rate from the WD.

Two of the more detailed models for the density and velocity structure of a CE in a nova system come from calculations by Livio et al. (1990) and Kley et al. (1995). Both groups modeled a common envelope appropriate for a post-outburst nova, assuming different mass loss rates for the wind impacting the secondary star. Both found the density of the CE to be concentrated near the binary orbital plane, and showed the envelope to rotate as it lost mass. The Kley model assumes a secondary star of 0.5 $M_\odot$ to be embedded in a wind of density n=$6 \times 10^{17}$ cm$^{-3}$, having a velocity of 425 km/s. The corresponding mass accretion rate onto the secondary star is $\approx 10^{-4}$ $M_\odot$/yr, which translates to an effective luminosity of order $10^3$ $L_\odot$. This result indicates a common envelope for the nova system to be viable energetically in accounting for a high post-outburst luminosity.

The Livio et al. models have an interior density of order $10^{16}$ cm$^{-3}$, and they find that spin-up of the CE occurs rapidly, in a matter of days, for a mass loss rate of order $10^{-6}$ $M_\odot$/yr, significantly less than the Kley group model. Insufficient to counteract the build-up of angular momentum in the envelope the lower mass loss rate leads to appreciable rotation that causes the frictional heating and luminosity of the envelope to become a less significant component of the ejecta emission.

Differences in the results of the Kley et al. and Livio et al. CE models may explain the unusual post-outburst behavior of V5856 Sgr. Its two most unusual features are its sustained luminosity after outburst and very narrow permitted emission line widths. The continued high luminosity is likely driven by nuclear reactions on the WD, which outburst models have shown will occur for certain input parameters (Yaron et al. 2005). The presence of a common envelope will depend on the masses of the stars, the size of the secondary orbit, and the rate of mass ejection from the WD. High mass loss rates following outburst lead to a more extensive CE that has a lower rotation rate, with emission having narrower line widths. Lower WD mass ejection rates lead to a CE with high angular momentum per until mass, producing broader line profiles. The unusual V5856 Sgr line profiles may be due to conditions for that system that create a substantial common envelope having low rotation, together with an expanding shell that emits the broad line component.

The presence of a narrow component, i.e., FWHM $\lesssim$ 60 km/s, that exhibits no radial velocity variation in any of the emission-lines, together with a significantly elevated brightness above pre-outburst levels may be the signature of a common envelope in classical novae. A number of novae have been observed to have narrow or sharp spike-like features in their emission lines, e.g., DE Cir/2003, YY Dor/2004, QY Mus/2008, KT Eri/2009, V2672 Oph/2009, LMC 2009a, and U Sco/2010, whose spectra have been obtained as part of the Walter et al. (2012) CTIO/SMARTS program. The majority of these objects are recurrent novae whose spectra exhibit radial velocity variations that are likely due to one of the stars, a re-established accretion disk, or polar jets (Munari, Mason, & Valisa 2014; Takeda & Diaz 2015). Thus, the presence of narrow line emission components themselves do not argue for common envelopes when they exhibit radial velocity variations. However, for those novae whose narrow components do not have detectable velocity variations and that remain unusually luminous some years after outburst, a common envelope is a plausible source of the emission.

Clear evidence for the existence of a common envelope must eventually come from using realistic values for the density and temperature of CE's to perform radiative transfer calculations that show the predicted emission-line spectrum can match the observations of novae that have unusual post-outburst characteristics. From the models of Livio et al. (1990) and Kley et al. (1995), common envelopes that are not in hydrostatic equilibrium have density scale-heights much larger than normally found above the photosphere of main sequence stars. This leads to a larger optically thin region that will produce more prominent emission lines relative to the continuum-forming region below it. Due to the non-spherical geometry of CE's in a binary system, orientation of the orbital plane to the line of sight is an important parameter in determining the characteristics of the emitted spectrum. Successful modeling of the spectrum emitted by a CE should constitute solid evidence for their existence in novae.

For V5856 Sgr, a post-outburst common envelope powered by nuclear reactions on the WD surface that drive mass loss may explain why (a) its emission lines show no perceptible radial velocity variations, (b) its narrow emission component has a small velocity width of FWHM∽50 km/s, (c) it has maintained such an unusually high UV and visible post-outburst luminosity, and (d) the nova is only marginally detected in X-rays. For novae with these characteristics serious consideration should be given to a common envelope as the primary source of the UV + visible radiation.

## 6. Summary

The spectrum of V5856 Sgr evolved rapidly after its discovery. In the weeks following outburst broad emission and absorption lines were observed that switched between emission and absorption spectra. Line widths indicate the ejection of mass having velocities of order $10^3$ km/s. The steady increase in brightness of the nova in the two weeks following discovery suggests continuous ejection was taking place over this interval. Expansion of the initial ejecta, likely to have been a ballistic ejection that formed a rapidly expanding shell, led to a decrease in its density to the point that broad, low ionization [N I] and [O I] forbidden lines appeared one month after outburst.

Spectra taken seven months after discovery revealed strong emission lines, with the H and He I profiles displaying prominent narrow emission components protruding above the broad base of the lines. At that time the level of ionization had increased, with [O III], [N II], and [O II] present at strengths comparable to that of H$\beta$ but having no narrow component detectable. The nova spectrum has maintained the same basic structure to the present time, with the relative strengths of the narrow/broad components of the permitted lines quite variable. The variability may be due to continuing activity on the WD, e.g., on-going nuclear reactions near its surface that provide both luminosity and mass loss for the common envelope.

The structure that is inferred for the radiating ejecta from its spectral evolution is that energy from the initial WD thermonuclear runaway produced a brief several-week period of mass loss that may also have triggered accompanying mass loss from the secondary star in the binary system, that may explain the apparent *thea* absorption lines. This initial burst of ejecta formed an expanding shell of gas that is the source of the broad emission lines in the spectrum. Prolonged nuclear reactions near the WD surface are subsequently driving a more continuous mass loss from the binary system that has kept the luminosity high and led to the formation of a common envelope.

Much of the direct energy that maintains the CE may derive from the orbital motion of the secondary star within the CE. The CE is not in hydrostatic equilibrium, so is losing mass. The CE's outer envelope is the source of the UV, visible, and IR continuum and narrow component of emission lines. The emission line characteristics constrain the CE line forming region to have densities in the range $10^8 - 10^{12}$ cm$^{-3}$, i.e., sufficiently high that collisional de-excitation inhibits forbidden line emission, and low enough that the O I 3d $^3$D$^o$ level does not populate the 3d $^5$D$^o$ level by collisional de-excitation that would produce a narrow emission component for O I $\lambda$9264 and $\lambda$7773. The prolonged high post-outburst brightness and the unusual spectral characteristics of V5856 Sgr are key attributes that can be produced by a common envelope formed after the outburst.

The authors wish to thank Dr. Nye Evans for his thorough review of this work, that was helpful in improving the presentation and interpretation of the data. FW acknowledges support from NSF grant

**Figure 1**

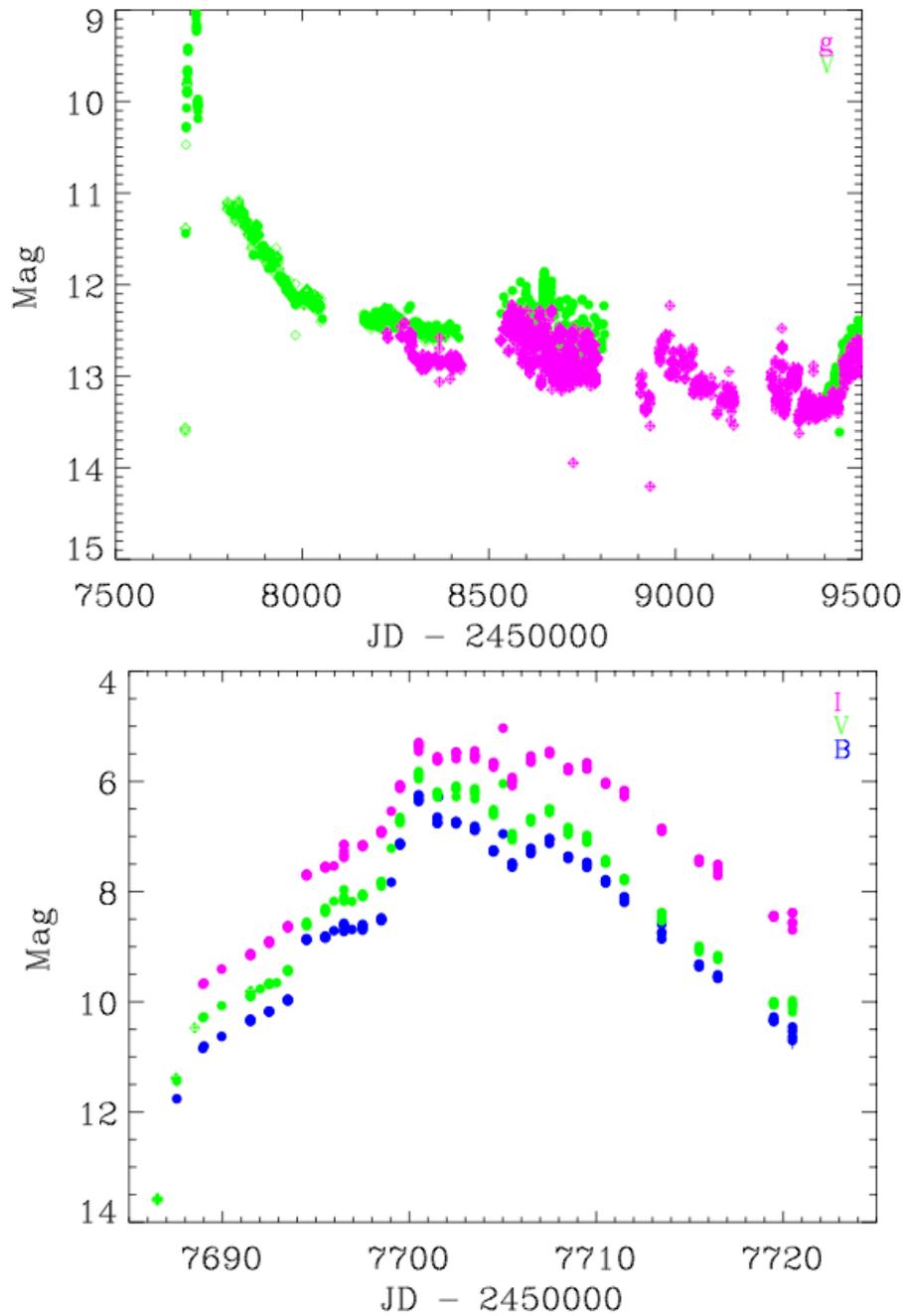

Figure 1 – The visible light curve of nova V5856 Sgr/2016 taken from photometric data. The top plot shows the flattening of the decline that leaves the visible brightness nine magnitudes above its pre-outburst level of I>22 (Munari et al. 2017), determined from pre-outburst ASASSN observations. The bottom curve shows a finer grid of dates that display the deliberate, slow steady rise and fall of the nova from its discovery, through visible maximum.

**Figure 2**

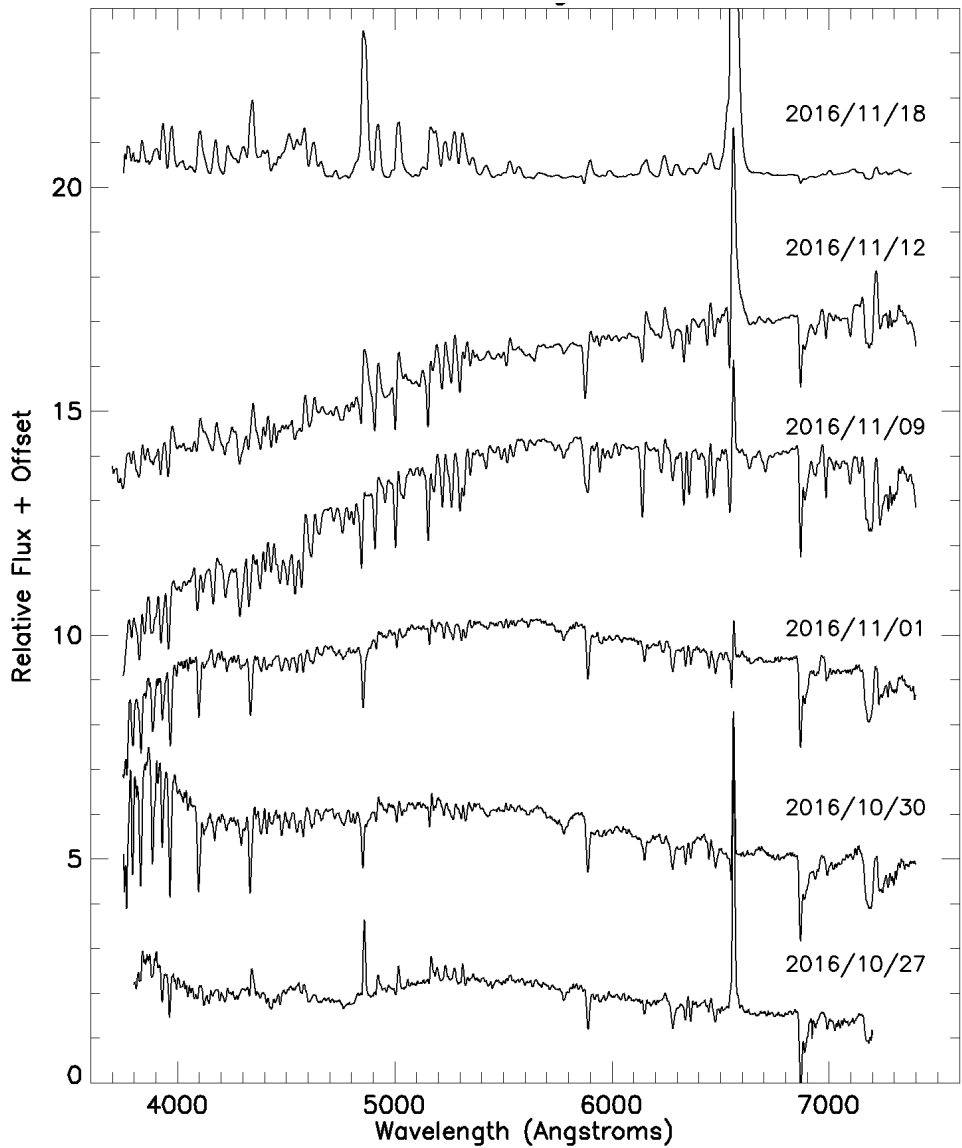

Figure 2 – A series of low-resolution spectra taken by Paul Luckas in the weeks following outburst. The initial spectrum was the basis for his determination of the nova as Fe II-class, from the prominence of Fe II multiplet 42. The nova quickly evolved to display absorption lines that became prominent at visual maximum. Shortly thereafter the spectrum transformed back into an emission spectrum similar to that it displayed at the time of discovery, although with broader lines reflecting higher gas velocities.

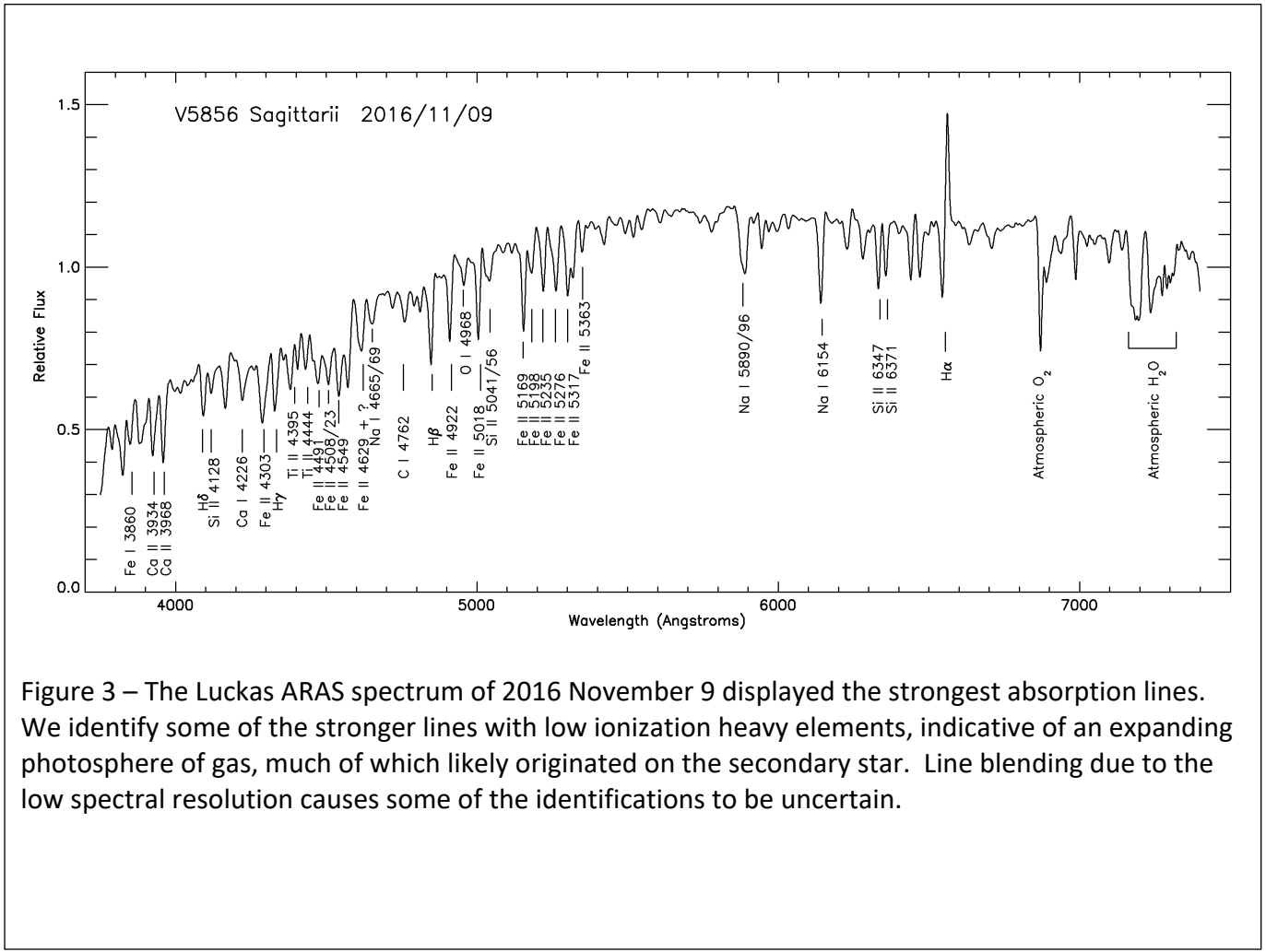

Figure 3 – The Luckas ARAS spectrum of 2016 November 9 displayed the strongest absorption lines. We identify some of the stronger lines with low ionization heavy elements, indicative of an expanding photosphere of gas, much of which likely originated on the secondary star. Line blending due to the low spectral resolution causes some of the identifications to be uncertain.

**Figure 4**

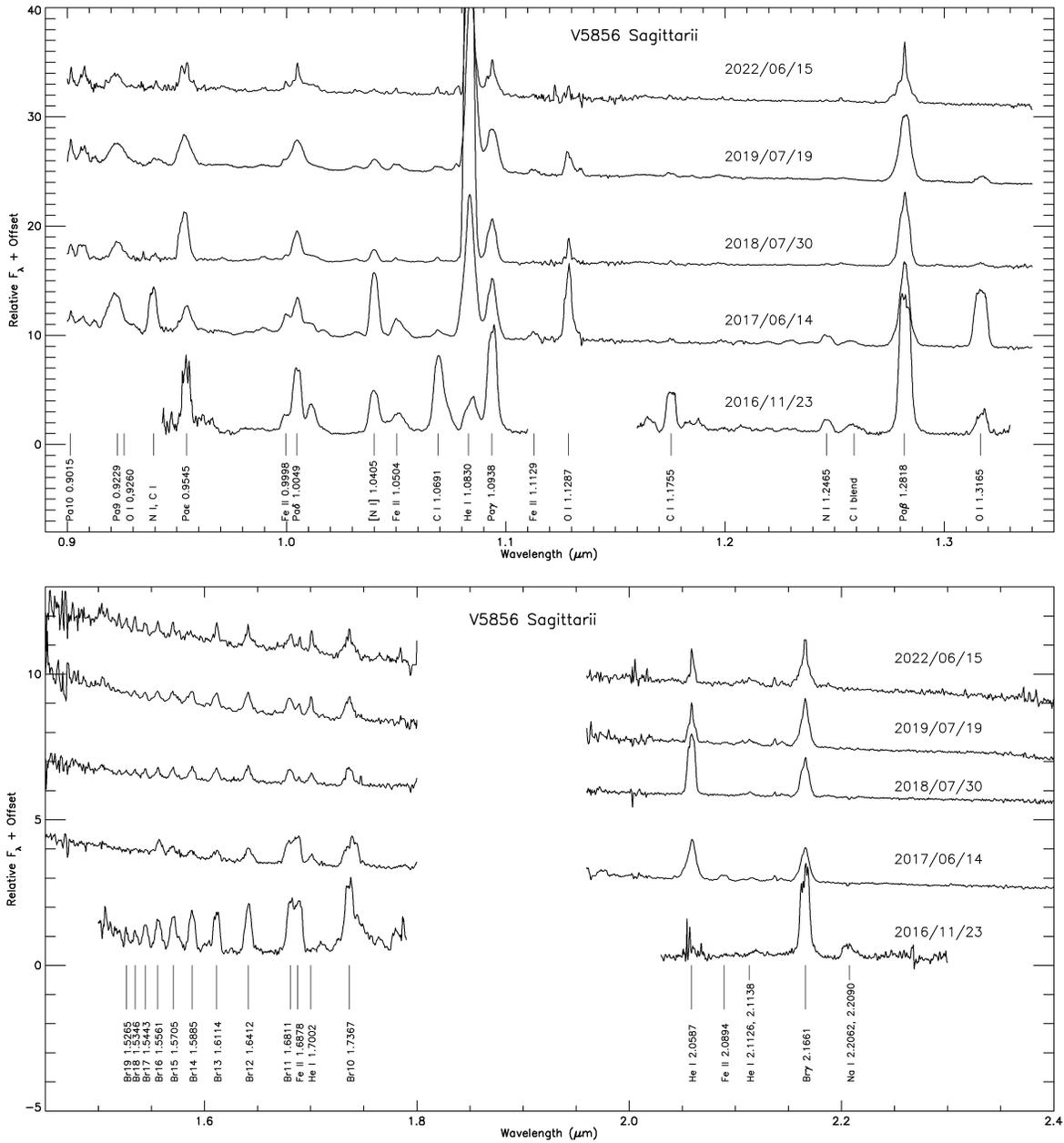

Figure 4 – The evolution of the infrared spectrum of the nova in the (a) far red + J band, and (b) H&K bands. Obtained with the Aerospace VNIRIS spectrograph, the 2016 observation was acquired using the Aerospace Corporation's 1m reflector, all subsequent measurements used the Lick Shane 3m telescope. Neutral CNO emission lines are notably strong immediately following outburst, with little evidence for higher ionization at later epochs.

**Figure 5**

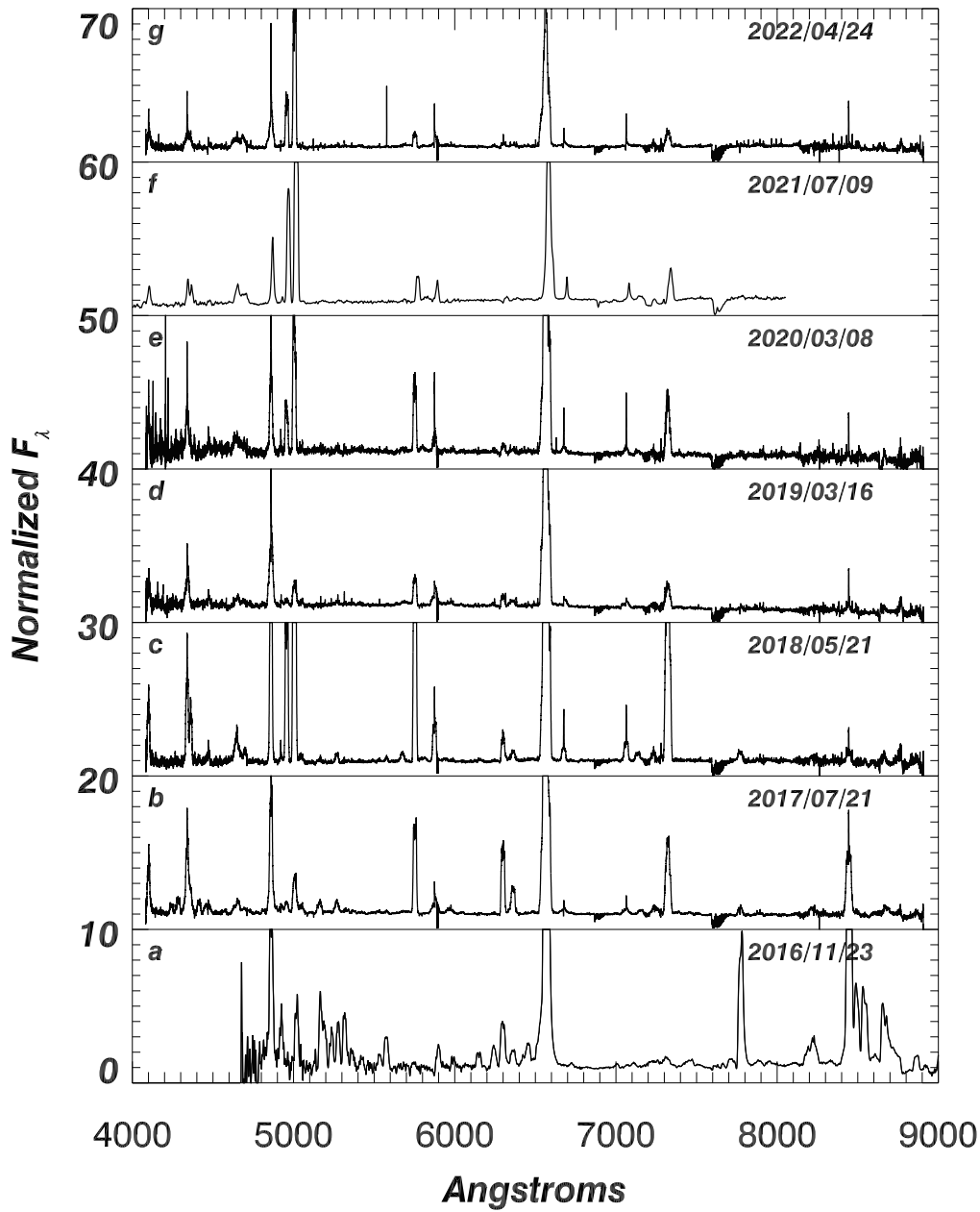

Figure 5 – Spectral evolution of V5856 Sgr in the visible over the 6-year period between outburst and the present time (2022). Data were obtained from (a) Aerospace 1m AEROTEL telescope with VNIRIS spectrograph at coude' focus; (b-e & g) CTIO/SMARTS + Chiron spectrograph, (f) Asiago 1.22m telescope + B&C Cassegrain spectrograph.

**Figure 6**

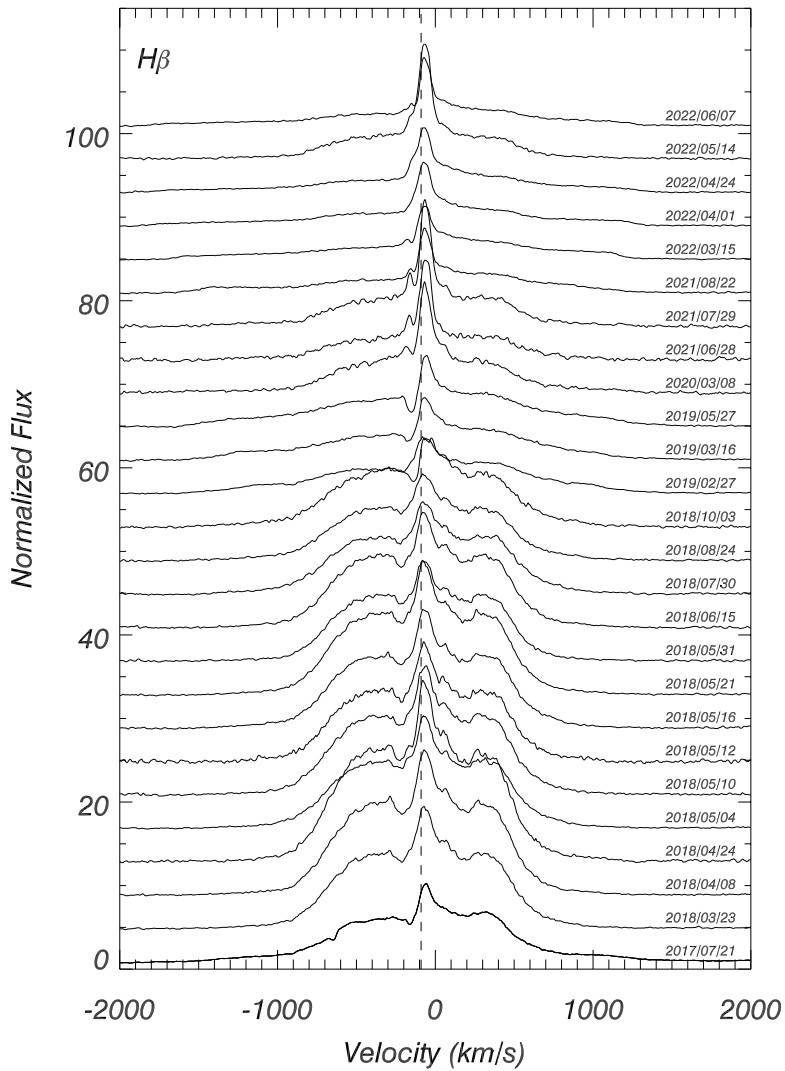

Fig. 6 – Chiron spectra of resolution 11 km/s show the evolution of Hβ over time. The flux in each plot is normalized to the continuum flux near −2,000 km/s. The line profile shows two clear components, one broad and the other narrow. Each dominate at different times, but both are always present. Only the narrow component clearly shows absorption, with its P Cygni profile indicating an expanding gas of order ≲100 km/s. The dashed vertical line represents the median velocity of the profile, corresponding to a radial velocity of the inhomogeneous emitting gas of −90 km/s.

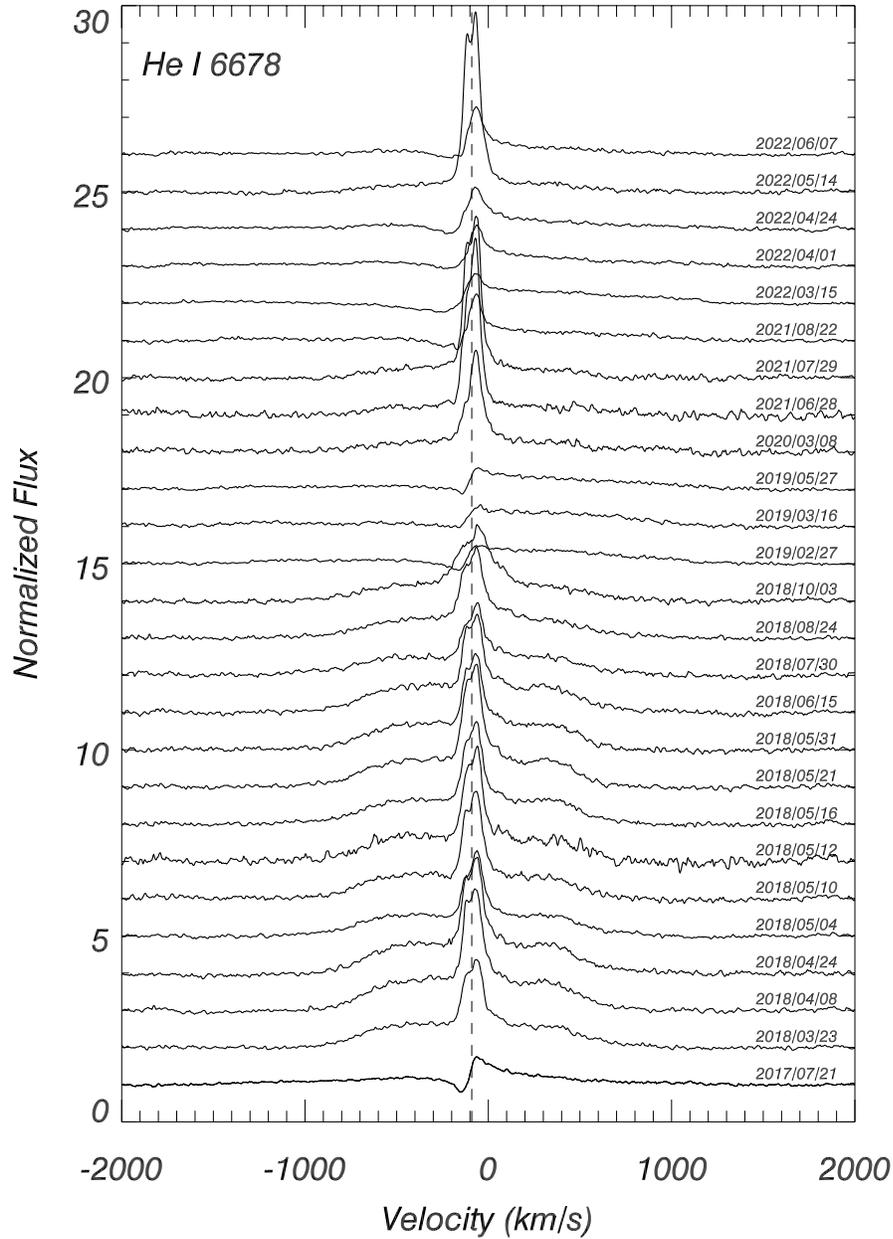

Fig. 7 - Chiron spectra of He I λ6678 showing its evolution. As in all plots in Figs. 6-11, the flux is normalized to the continuum flux near -2,000 km/s. Compared to Hβ the λ6678 broad emission component is less prominent relative to its narrow component.

# Figure 8

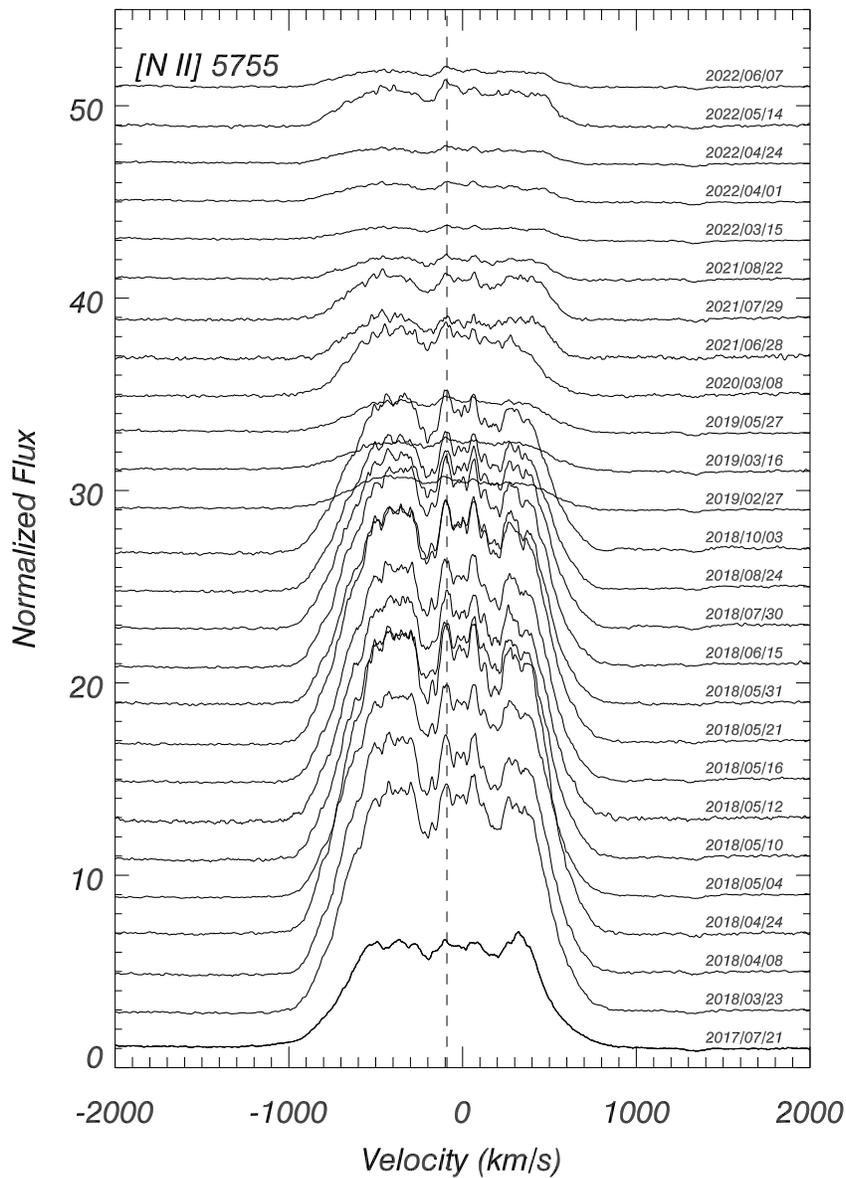

Fig. 8 - Chiron spectra showing the evolution of [N II] λ5755. The individual spectra are plotted with respect to the normalized continuum flux, and show considerable variation in the strength of [N II] over time relative to the continuum, whose brightness has not varied significantly in the visible over the past six years. The continuum brightness changes by a factor of more than 2 among the different dates, and the profile consists entirely of the broad emission component.

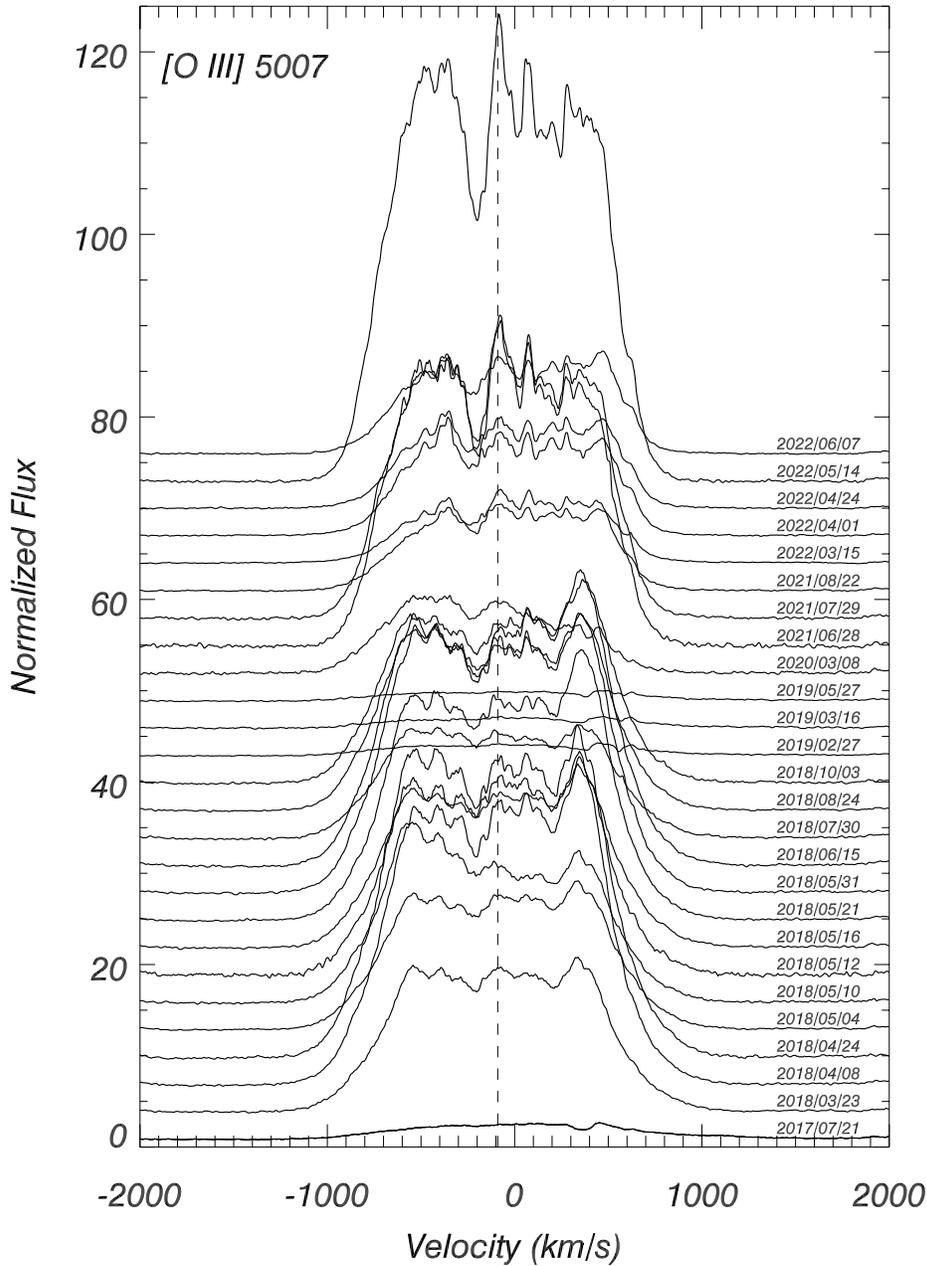

Fig. 9 - Chiron spectra of [O III] λ5007 at high resolution of 0.18 Å (11 km/s).  After the initial optically thick fireball stage, where permitted lines dominate the spectrum, the [O III] emission becomes prominent.  The emission peak at ~400 km/s at early epochs is due to He I λ5015 emission.

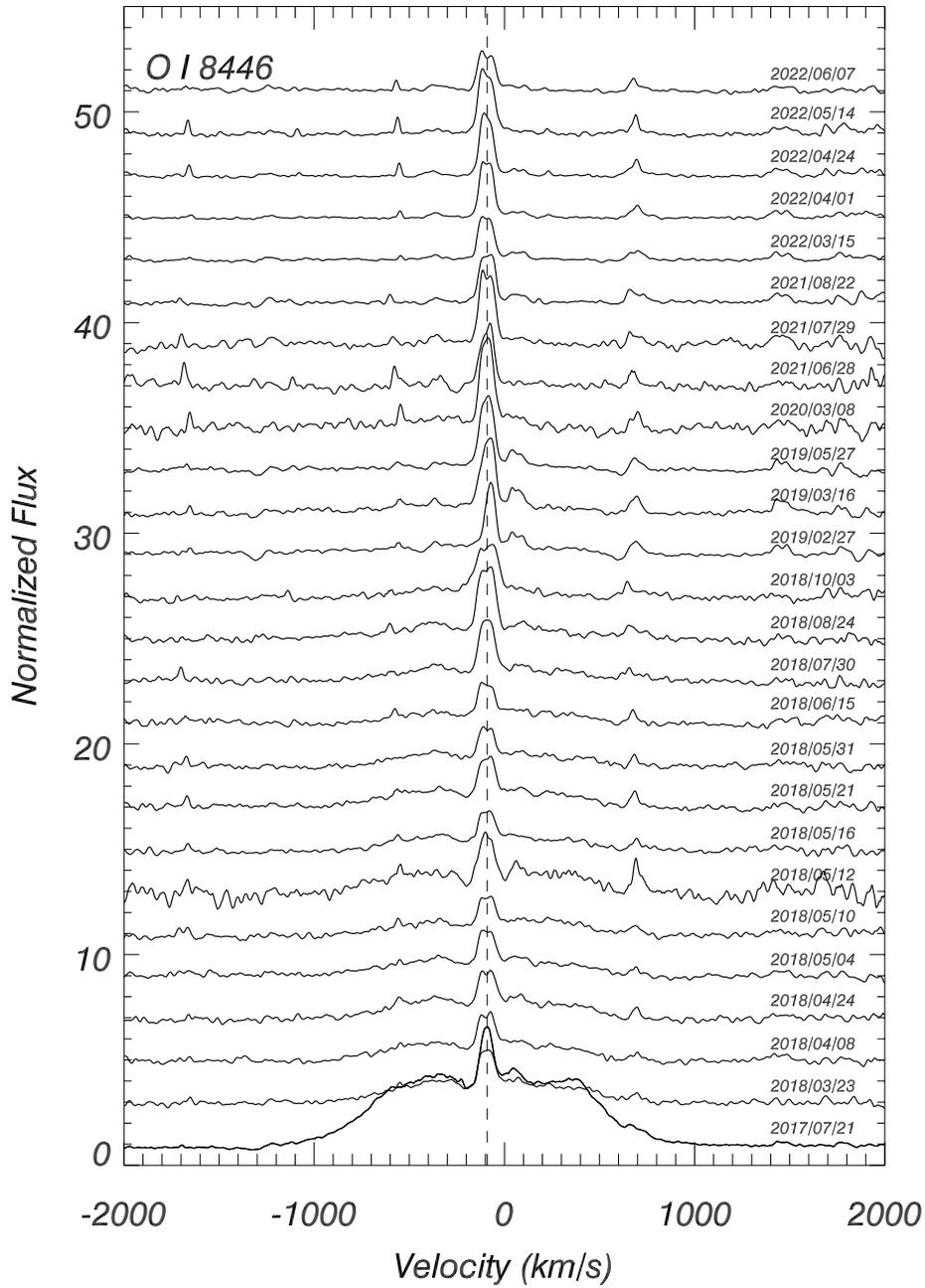

Fig. 10 – High resolution Chiron spectra of the O I λ8446 line. The narrow component dominates the emission after two years following the outburst.

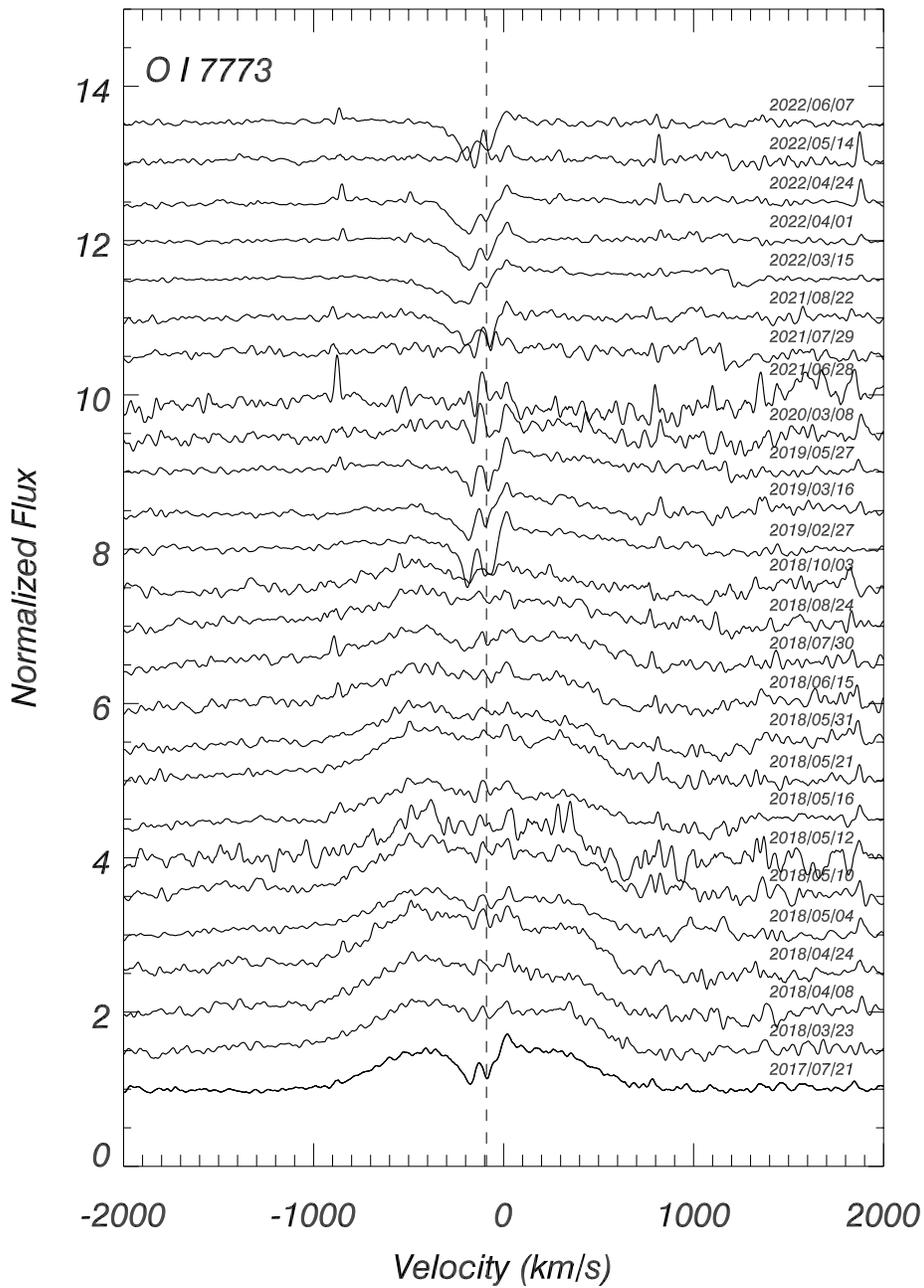

Fig. 11 - Chiron spectra of the O I λ7773 multiplet at high resolution of 11 km/s. Resolved absorption from intervening gas is frequently present, but no narrow emission component is detected at any time. As with the O I λ8446 line, broad emission is present for the early years following outburst, but fades into the continuum thereafter.

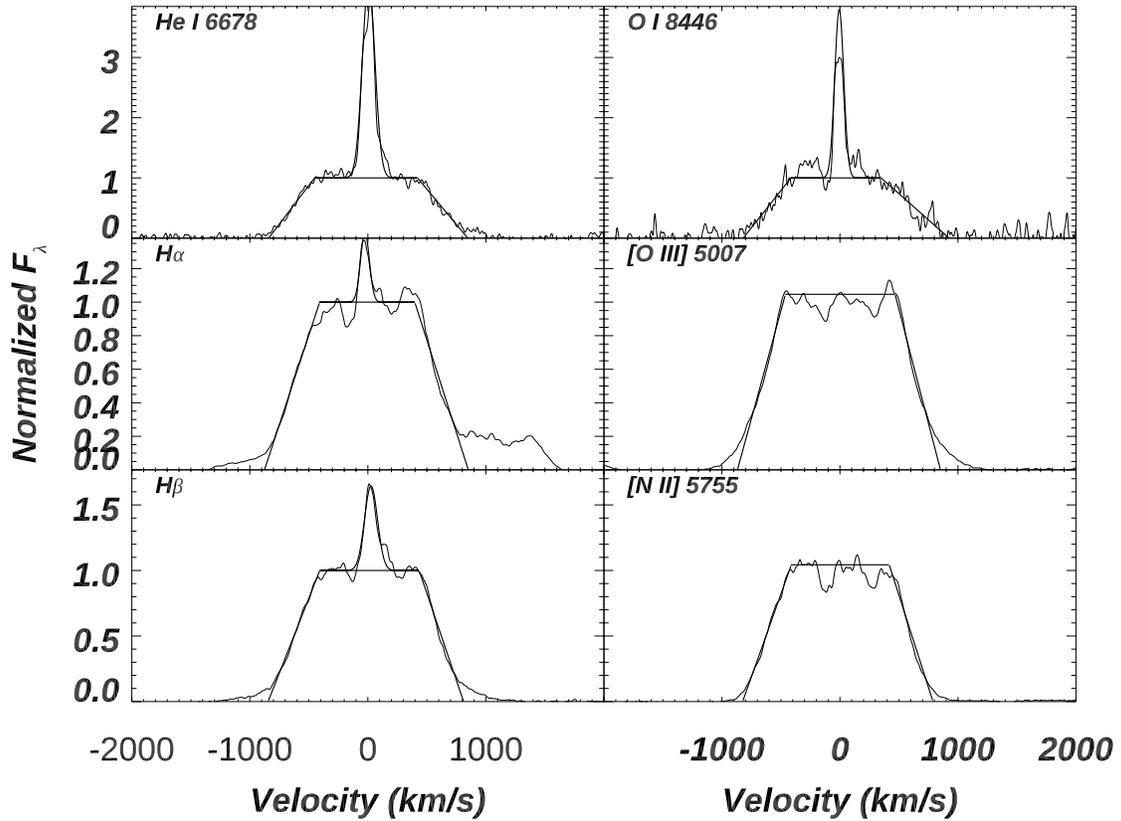

Fig. 12 – Deconvolution of the narrow and broad components of the stronger permitted and forbidden emission lines in the 2018 March 23 spectrum. The broad component can be fitted adequately with a trapezoid, whereas the narrow component, whose profile is influenced by blueward absorption, is approximately Gaussian.

**Figure 13**

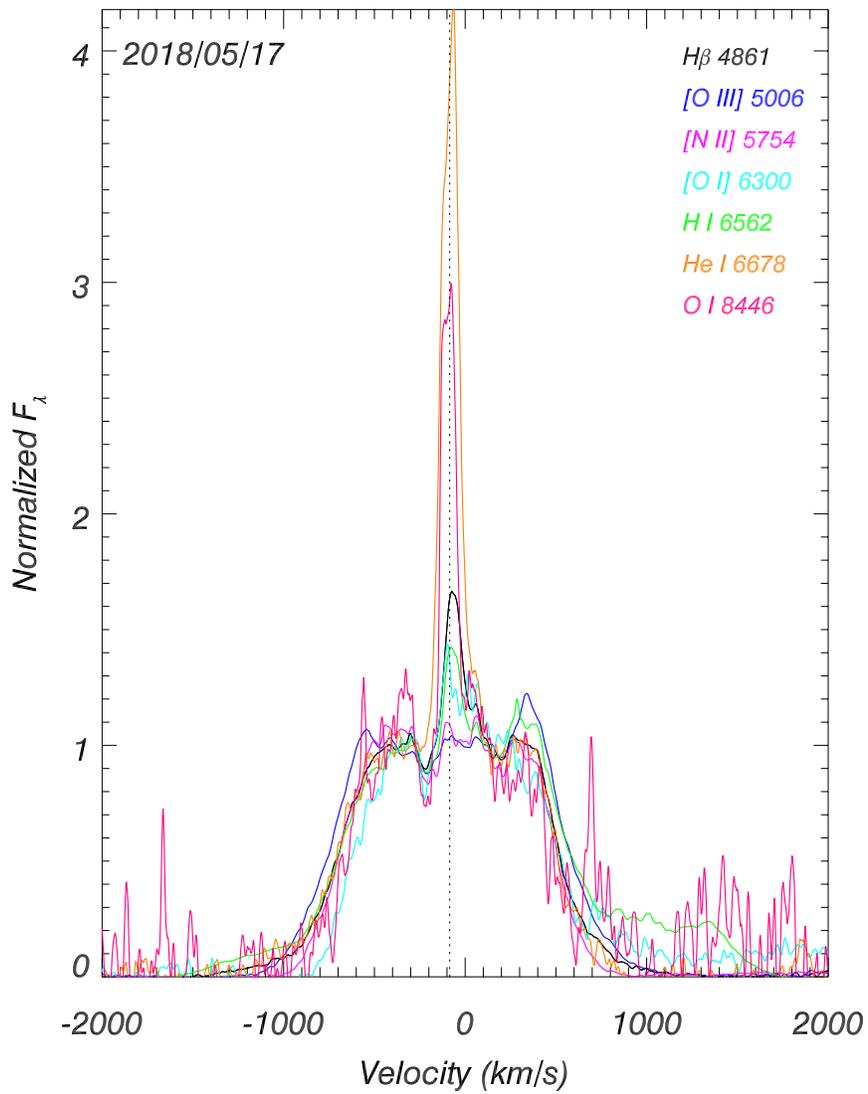

Fig. 13 – A direct comparison of profiles of the stronger emission lines during post-outburst decline that show the shapes and widths of the different types of lines. The profiles are normalized so the flat peaks of the broad components have the same flux. The forbidden lines and the Balmer and He I lines have very similar widths.